# Different strategies for GaN-MoS$_2$ and GaN-WS$_2$ core-shell nanowire growth

*Edgars Butanovs[1,2]\*, Kevon Kadiwala[1], Aleksejs Gopejenko[1], Dmitry Bocharov[1], Sergei Piskunov[1], Boris Polyakov[1]*

[1]Institute of Solid State Physics, University of Latvia, Kengaraga street 8, Riga, Latvia, LV-1063

[2]Institute of Technology, University of Tartu, Nooruse 1, 50411 Tartu, Estonia

\*e-mail: edgars.butanovs@cfi.lu.lv

**Abstract**

One-dimensional (1D) nanostructures – nanowires (NWs) – exhibit promising properties for integration in different types of functional devices. Their properties can be enhanced even further or tuned for a specific application by combining different promising materials, such as layered van der Waals materials and conventional semiconductors, into 1D-1D core-shell heterostructures. In this work, we demonstrated growth of GaN-MoS$_2$ and GaN-WS$_2$ core-shell NWs via two different methods: (1) two-step process of sputter-deposition of a sacrificial transition metal oxide coating on GaN NWs followed by sulfurization; (2) pulsed laser deposition of few-layer MoS$_2$ or WS$_2$ on GaN NWs from the respective material targets. As-prepared nanostructures were characterized via scanning and transmission electron microscopies, X-ray diffraction, micro-Raman spectroscopy and X-ray photoelectron spectroscopy. High crystalline quality core-shell NW heterostructures with few-layer MoS$_2$ and WS$_2$ shells can be prepared via both routes. The experimental results were supported by theoretical electronic structure calculations, which demonstrated the potential of the synthesised core-shell NW heterostructures as photocatalysts for efficient hydrogen production from water.

# 1. Introduction

Layered two-dimensional (2D) van der Waals (vdW) materials have been a hot research topic since the ground-breaking experiments on graphene in 2004 [1]. 2D vdW materials, such as graphene, hexagonal boron nitride, transition metal dichalcogenides (TMDs), have been shown to exhibit promising electrical, optical and mechanical properties when their thickness is reduced to several atomic layers [2]. TMDs have a general chemical formula $MeX_2$, where Me is a periodic table Group 4 – 7 transition metal and X is S, Se or Te. Most commonly studied TMDs are typically semiconductors with indirect-to-direct bandgap transition from bulk to monolayer [3]. For example, $MoS_2$ bandgap shifts from 1.2 to 1.9 eV when reduced to monolayer, while $WS_2$ bandgap shifts from 1.3 to 2.1 eV [4]. TMDs are now being investigated for potential applications in energy [5,6], sensors [7], electronics and optoelectronics [4].

On the other hand, one-dimensional (1D) nanostructures – nanowires (NWs) – have been extensively studied for more than two decades and exhibit promising properties for integration in different types of functional devices [8,9]. By preparing hybrid material NWs, for example, various 1D-2D or axial and radial 1D heterostructures, their characteristics can be improved even further [10,11]. One type of such hybrid nanostructures are *core-shell* NWs, which are promising due to possible modification of NW electronic surface states [12] and even further increase of already high surface-to-volume ratio [13]. A potential way to effectively increase the NW surface is to grow layered vdW materials around the NW, since they exhibit weak interlayer bonding [14,15]. Many energy applications, e.g., photo- and electrocatalytic hydrogen evolution reactions (HER), supercapacitors and storage, require materials with large effective surface and high concentration of active sites, therefore, NW and layered material heterostructures are promising candidates [16–23]. Furthermore, the NW core can also be a high-mobility charge carrier migration channel, if the right materials are selected [24].

Gallium nitride (GaN) nanoscale heterostructures with mono- or few-layer $MoS_2$ and $WS_2$ have recently been shown to exhibit type-II band alignment, which is suitable for optoelectronics and photoelectrocatalysis applications [25,26]. GaN is an established material for optoelectronics due to its direct bandgap ($E_g \sim 3.45$ eV) and high electron mobility in the form of thin films and NWs [27], while the thickness-dependent bandgap of layered 2D materials gives tunability for desired applications [28]. In the case of GaN-$MoS_2$ heterostructures, there have been several experimental and theoretical studies of 2D-2D (thin films or thin crystals) heterojunctions for applications as photodetectors [29–31], sensors [32] and catalysts for water splitting [33,34]. Furthermore, 1D-2D GaN-$MoS_2$ heterostructures have been demonstrated in electronic and optoelectronic devices [35–37]. As for GaN-$WS_2$, a combination of these materials shows promise in optoelectronics [38] and photocatalysis [25], since $WS_2$ can also be an efficient catalyst [39]. Regarding 1D-1D *core-shell* nanostructures, growth of $MoS_2$-wrapped GaN NWs has recently been demonstrated by *Ji et al.* via direct deposition of $MoS_2$ on GaN NWs with magnetron sputtering [40] and GaN-$MoS_x$ *core-shell* NWs by *Zhou et al.* via electrodeposition [41]. The few previous studies clearly demonstrate the potential of such heterostructures; however, the development of other versatile fabrication methods is still needed for further progress.

In this work, we demonstrated growth of GaN-$MoS_2$ and GaN-$WS_2$ *core-shell* NWs via two different methods: (1) two-step process of sputter-deposition of a sacrificial respective transition metal oxide coating on GaN NWs followed by sulfurization; (2) pulsed laser deposition of few-layer TMDs on GaN NWs from $MeS_2$ targets. Characterization results show that high crystalline quality *core-shell* NW heterostructures can be prepared through both routes. To theoretically support the experimental results, the electronic structure of a model reproducing structural properties *core-shell* NWs was investigated. The model used here differs from numerous models of heterostructures where (0001) $MoS_2$/$WS_2$ surface possessing the hexagonal symmetry is deposited on the GaN (0001) surface [25,28,42–44], which also exhibits hexagonal symmetry. The model previously proposed for structurally similar $ZnO$/$WS_2$ heterointerface [45] was used in this work. In this model GaN ($1\bar{1}00$) surface of GaN substrate is stacked with the outer atomic layer of $MoS_2$/$WS_2$ (0001). In contrast to the $ZnO$/$WS_2$ model, in which bridging atoms between the surfaces were used to stack both interfaces, here the stacking occurred directly due to a better match between the positions of the surface atoms of sulfur and gallium. This model corresponds well to the situation of TMD shell growth on a gallium nitride [0001]-oriented NW core, which is important for the correct description of nanowire electronic properties. Based on our calculations, GaN-$MoS_2$ and GaN-$WS_2$ *core-shell* NWs might be promising candidates for further investigation as photocatalysts for efficient hydrogen production from water.

**2. Experimental details**

GaN-$MeX_2$ *core-shell* NW preparation methods demonstrated in this work are schematically depicted in *Fig. 1*. GaN NWs were synthesized via atmospheric pressure chemical vapour transport method in a horizontal quartz tube reactor. 2 g metallic Ga (99.999%, Alfa Aeasar) was loaded in a ceramic boat and placed in the centre of the quartz tube, oxidized silicon wafers $SiO_2$/Si(100) (*Semiconductor Wafer, Inc.*) coated with spherical Au nanoparticles (NPs, *Alfa Aesar*, water suspension, 100 nm diameter) were placed downstream in a lower temperature region. Au NPs were used as a catalyst for the vapour-liquid-solid (VLS) mechanism. The reactor was heated to 940°C under a flow of carrier gas mixture Ar/H$_2$-35%, then gaseous NH$_3$ flow in 1:1 ratio to the carrier gas was introduced and maintained for 30 minutes for the gas-phase reaction and NW growth, followed by natural cooling to the room temperature under Ar/H$_2$ flow. As a result, 5-20 μm long GaN NWs were produced on the SiO$_2$/Si substrate. Characterization data for the as-grown pure GaN NWs is shown in *Fig. S1*.

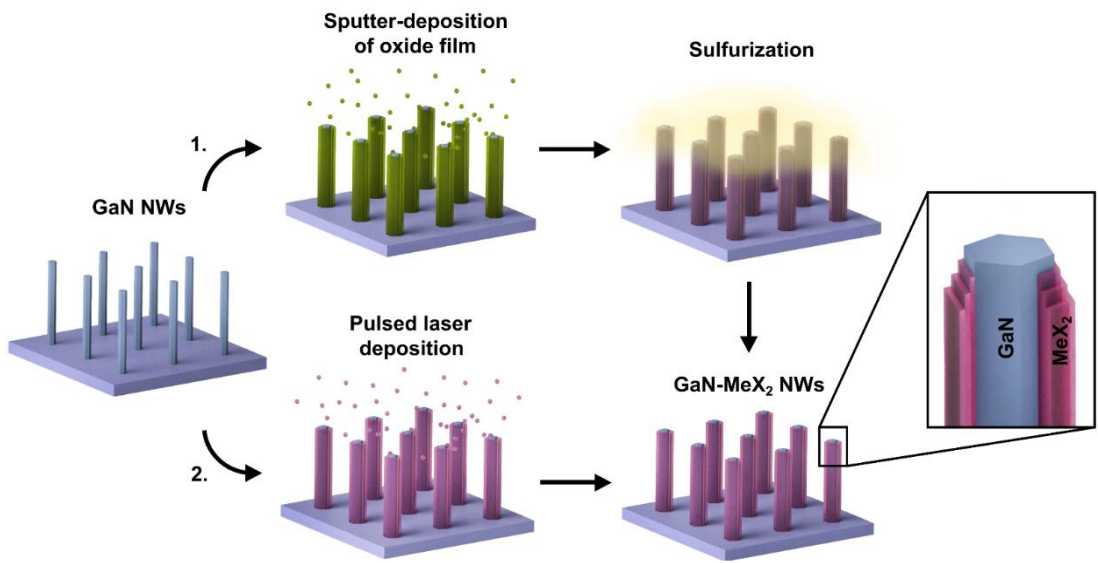

**Figure 1**. A schematic of both demonstrated GaN-MeX$_2$ core-shell NW preparation methods on Si/SiO$_2$ substrates: (1) two-step method, which includes sulfurization of pre-deposited metal oxide coating; and (2) direct deposition of MoS$_2$ or WS$_2$ with pulsed laser deposition.

Few-layers of MoS$_2$ and WS$_2$ on GaN NWs were obtained with two different routes. The first route consists of two steps – deposition of amorphous MoO$_3$ and WO$_3$ coating on GaN NWs via reactive DC magnetron sputtering of a metallic target in a mixed Ar/O$_2$ atmosphere, followed by subsequent sulfurization of the samples in a quartz tube reactor at high temperatures. The optimal sacrificial precursor film thickness (on a flat substrate) was found to be 30 nm and 40 nm for MoO$_3$ and WO$_3$, respectively. Optimal sulfurization temperature was 750°C for MoS$_2$ and 800°C for $WS_2$ coatings. The second route was pulsed laser deposition (PLD) from stoichiometric $MoS_2$ and $WS_2$ targets. 500 mJ 248 nm KrF laser beam was used for target ablation at 10 Hz repetition frequency and $10^{-5}$ Torr background pressure. A few-layer $MoS_2$ coating was obtained with 1500 pulses at 600°C substrate temperature, and $WS_2$ coating with 3000 pulses at 650°C substrate temperature.

As-grown NW morphology was characterized using a scanning electron microscope (SEM, Lyra, Tescan), while their inner crystalline structure was studied using a transmission electron microscope (TEM, Tecnai GF20, FEI) operated at a 200 kV accelerating voltage. NW phase composition was studied by X-ray diffraction (XRD) using Rigaku MiniFlex 600 X-ray powder diffractometer with Bragg-Brentano θ-2θ geometry and the 600W Cu anode (Cu Kα radiation, λ = 1.5406 Å) X-ray tube. Micro-Raman spectroscopy measurements were performed using a TriVista 777 confocal Raman system (Princeton Instruments, 750 mm focal length, 1800 lines/mm grating) equipped with an upright Olympus microscope with Olympus UIS2 MPlanN 100x/0.90 objective, a continuous-wave single-frequency diode-pumped laser Cobolt Samba 150 (λ=532 nm) and Andor iDus DV420A-OE CCD camera. XPS measurements were performed using an X-ray photoelectron spectrometer ESCALAB Xi (ThermoFisher), and XPSPEAK41 software was used for peak fitting.

**3. Theoretical background and computational model**

CRYSTAL17 computer code [46,47], which employs Gaussian-type functions centred on atomic nuclei as the basis sets (BS) for an expansion of the crystalline orbitals has been used to perform the DFT calculations of GaN-$MoS_2$ heterostructures implementing hybrid HSE06 exchange-correlation functional. Heyd-Scuseria-Ernzerhof hybrid exchange-correlation functional (HSE06) [48], which uses a screened hybrid functional and includes the exact nonlocal Fock exchange, has been used to perform the calculations of GaN, $MoS_2$, and $WS_2$ structures as well as $GaN/MoS_2$ and $GaN/WS_2$ heterostructures. Ga_86-4111d41G basis set [49] has been used for gallium atoms, N_6-31d1G basis set has been adopted for nitrogen atoms calculations [50], W_cora_1996 basis set [51] has been chosen for tungsten atoms, Mo_SC_HAYWSC-311(d31) Hay Wadt small core basis set [52] has been used to describe molybdenum atoms, and S_86-311G basis set taken from [53] has been used for sulphur atoms. To perform the calculations, the Brillouin zone has been sampled by 8×8×8 Pack-Monkhorst net [54] resulting in 75 *k*-points total. Empirical Grimme correction [55] is not used in the calculations.

In this work, we calculated the separate slabs of hexagonal 2H phase $WS_2$ (0001) and $MoS_2$ (0001) with the thickness from 1 to 3 monolayers of dichalcogenide, GaN ($1\bar{1}00$) slab with the thickness of 24 atomic layers, as well as the heterostructure formed by the 24 atomic-layer GaN slab on which 1-3 layers of $WS_2$ (0001) or $MoS_2$ (0001) were deposited. Each monolayer of dichalcogenide in the slab is formed by three atomic planes where the central plane contains only Me atoms and the outer planes contain S atoms. Neighbouring atomic monolayers of dichalcogenide consist of the inverse arrangements on the atoms. Three projections of $WS_2$ surface and GaN surface are shown in *Figs. S4* and *S5*, correspondingly. The heterostructures formed by GaN substrate with deposited $WS_2/MoS_2$ layers are shown in *Figs. S6* and *S7*.

**4. Results and discussion**

As-grown individual *core-shell* NW morphology and inner structure was characterized using TEM (see *Fig. 2*), while SEM was used to confirm that the NWs maintain their length after the shell deposition procedures, as can be seen in *Fig. S2*. Lower magnification TEM images show the typical coating morphologies obtained via each $MoS_2$ and $WS_2$ synthesis method. While with PLD approach and pre-deposited $WO_3$ sulfurization (*Fig 2. (c,e,g)*) it is possible to obtain a smooth and uniform shell around NWs, $MoO_3$ sulfurization at optimal conversion and crystallization temperature gives non-uniform island-like coating (see *Fig. 2(a)*). Uniform shell is usually paramount for good electrical properties; however, the TMD-NW hybrid nanostructures with increased surface roughness are excellent for various energy applications [56], since the TMD in this configuration provides a lot of open active edge sites, while the NW core keeps its highly crystalline structure. At higher TEM resolution (*Fig. 2(b,d,f,h)*) the inner crystalline structure of the nanostructures is revealed. The layers of synthesised $MeX_2$ (each consisting of X-Me-X atomic planes) shells can be distinguished as parallel black and white lines along the surface of NWs. The typical thickness of the uniform coatings varies from 4 to 10 monolayers, however, with some limitations it is possible to achieve thickness control including single monolayer deposition (see *Fig. S3*). Interplanar distance values (d-spacings) were measured to be around 6.4 – 6.6 Å for $MoS_2$ and 6.2 – 6.5 Å for $WS_2$, which are in good agreement with the lattice parameters for bulk $MoS_2$ (a = 6.16 Å, ICDD-PDF #37-1492) and $WS_2$ (a = 6.18 Å, ICDD-PDF #08-0237), respectively. Furthermore, the single-crystalline nature of the GaN NW core is clearly visible; the measured interplanar distance is 2.7 – 2.8 Å, closely matching hexagonal GaN (a = 2.76 Å, ICDD-PDF #50-0792). The TEM measurements show the high crystalline quality of the prepared *core-shell* NWs.

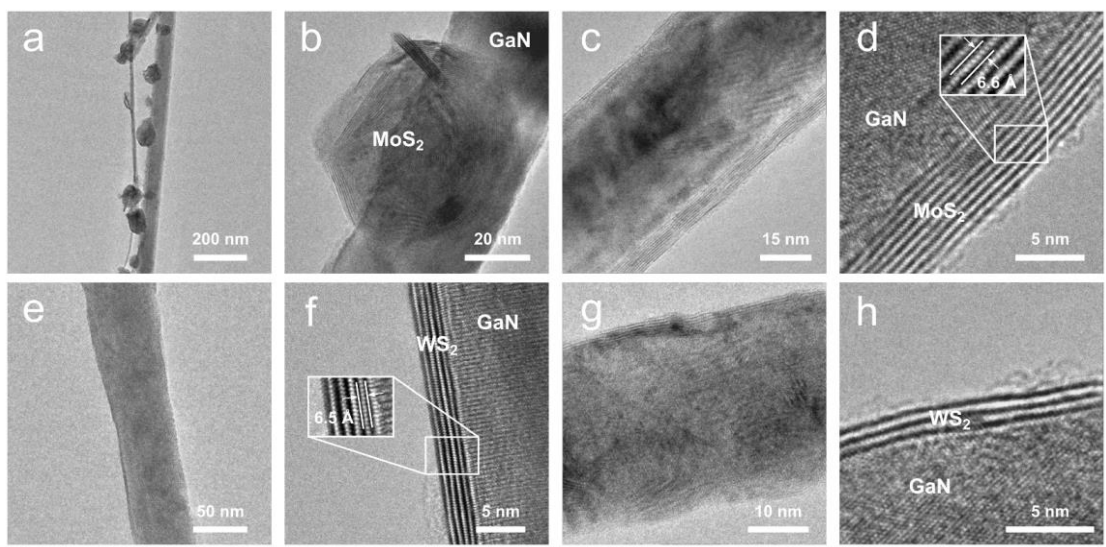

**Figure 2**. Transmission electron microscope images at different magnifications of individual (a-b) GaN-$MoS_2$ NW prepared via $MoO_3$ coating sulfurization, (c-d) GaN-$MoS_2$ NW prepared via pulsed laser deposition, (e-f) GaN-$WS_2$ NW prepared via $WO_3$ coating sulfurization, (g-h) GaN-$WS_2$ NW prepared via pulsed laser deposition; the insets show the measured d-spacings.

To confirm the presence of the respective phases in the as-grown *core-shell* NW samples, XRD measurements were performed on the NW arrays on the $Si(100)/SiO_2$ substrates (see *Fig. 3(a,c)*). Here and later on, for the sake of clarity, measurement results are shown only for the nanostructures prepared by the two-step sacrificial coating method, since the PLD samples gave qualitatively identical results. Both XRD patterns for each sample exhibit Bragg peaks of the desired crystalline phases: hexagonal GaN NWs (ICDD-PDF #50-0792), $MoS_2$ shell (ICDD-PDF #37-1492) and $WS_2$ shell (ICDD-PDF #08-0237), supporting the TEM measurements. Bragg peak at $2\theta \approx 33°$ can be attributed to the Si(100) substrate (forbidden Si(200) reflection). Peaks at 38.1° and 44.3° belong to the Au NPs used for the VLS growth (ICDD-PDF #04-0784). Furthermore, room-temperature micro-Raman spectroscopy was used to confirm the presence of the $WS_2$ and $MoS_2$ layers in the as-prepared nanostructures. The measured Raman spectrum of GaN-$MoS_2$ *core-shell* NW contains the in-plane $E^1_{2g}$ mode at 383 cm$^{-1}$ and the out-of-plane $A_{1g}$ mode at 408 cm$^{-1}$ (see *Fig. 3(b)*), thus confirming the formation of $MoS_2$ on GaN NWs [57]. Similarly, in the obtained GaN-$WS_2$ NW spectra, two bands at 352 cm$^{-1}$ and 420 cm$^{-1}$ were measured and attributed to the $E^1_{2g}$ and $A_{1g}$ modes in $WS_2$ [45], respectively (*Fig. 3(d)*).

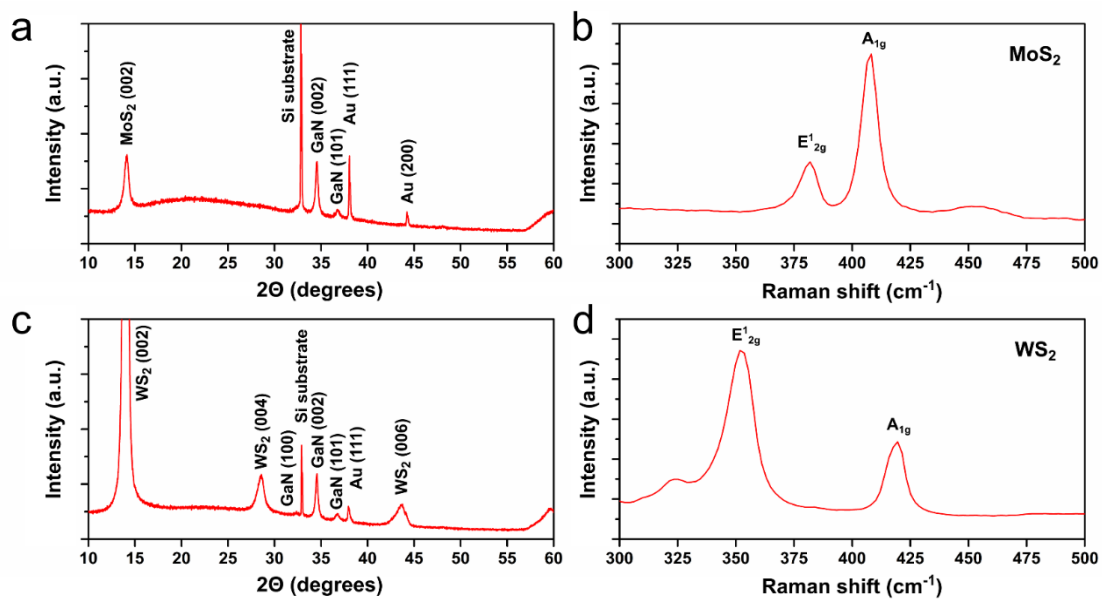

**Figure 3**. (a) X-ray diffraction and (b) micro-Raman spectrum of GaN-$MoS_2$ NW arrays on a $Si/SiO_2$ substrate; (c) X-ray diffraction and (d) micro-Raman spectrum of GaN-$WS_2$ NW arrays on a $Si/SiO_2$ substrate.

An XPS analysis was performed in order to verify the chemical composition of the *core-shell* NW arrays on the Si(100)/$SiO_2$ substrates (see *Fig. 4*). High-resolution spectra of Mo 3d, W 4f, S 2p, Ga 3d and N 1s peaks were acquired and calibrated relative to adventitious C 1s peak at 285 eV binding energy. Ga and N characteristic peaks for all samples confirmed the presence of GaN chemical state as was expected from GaN NWs: Ga 3d scan consists of Ga-N peak at 20.2 eV together with O 2s contribution at around 24.0 eV, while N 1s scan shows Ga-N peak at 397.8 eV and strong overlapping with Ga LMM peak. For both – GaN-MoS$_2$ and GaN-WS$_2$ – samples S 2p peaks consist of spin-orbit doublets ($\Delta_{3/2-1/2}$=1.2 eV) with 2p$_{3/2}$ peak measured at 161.8 eV, which is consistent with MoS$_2$ and WS$_2$ compounds. In the case of GaN-MoS$_2$ NWs, Mo 3d peak shows spin-orbit splitting ($\Delta_{5/2-3/2}$=3.2 eV) with Mo 3d$_{5/2}$ component being at 229.0 eV, corresponding to the MoS$_2$ chemical state [41]. Overlapping S 2s peak at 226.2 eV can also be observed in the scan. As for GaN-WS$_2$ NWs, the characteristic W 4f$_{7/2}$ peak position for WS$_2$ chemical state was measured at 32.0 eV ($\Delta_{7/2-5/2}$=2.2 eV) [39].

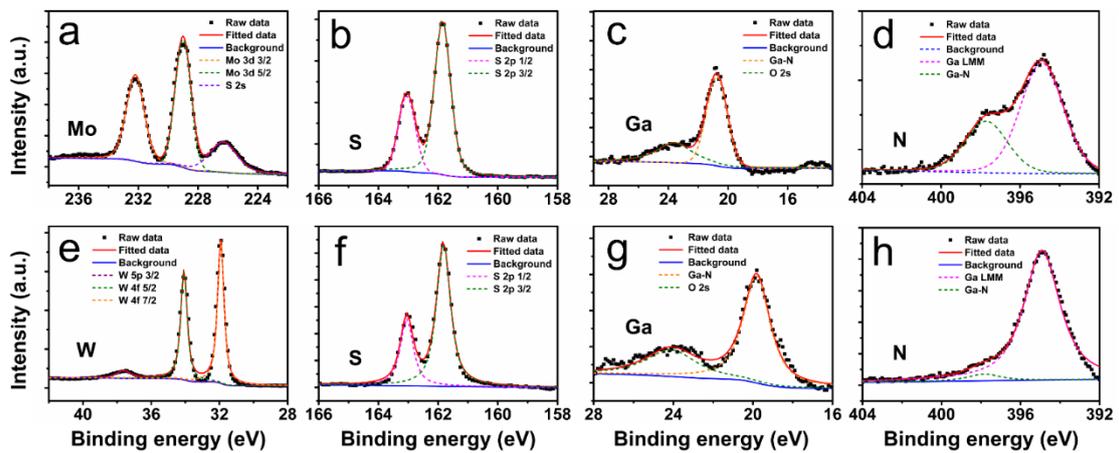

**Figure 4**. High-resolution XPS core-level spectra and peak fits of GaN-$MoS_2$ core-shell NWs for (a) Mo 3d, (b) S 2p, (c) Ga 3d and (d) N 1s. High-resolution XPS core-level spectra and peak fits of GaN-$WS_2$ core-shell NWs for (a) W 4f, (b) S 2p, (c) Ga 3d and (d) N 1s.

Within the framework of this study the calculations of GaN, WS$_2$ and MoS$_2$ bulk crystals, free-standing GaN, WS$_2$ and MoS$_2$ 2D slabs, as well as WS$_2$@GaN and MoS$_2$@GaN 2D nanoheterointerfaces have been performed. The calculated lattice constants of the GaN bulk are $a$=3.189 Å and $c$=5.162 Å, which are in a good agreement with experimental data $a$=3.189Å and $c$=5.186 Å [doi.org/10.1016/S0022-0248(96)00341-7]. The distance between the Ga atoms in the middle of the slab is 3.17Å, while the relaxation causes noticeable distortion of the atomic positions at termination edges of the slab, where the distance between the closest Ga atoms varies from 2.93 Å to 3.24 Å. The similar atomic shifts are calculated for the distances between Ga and N atoms, as well as between N atoms. The distance between Ga and N atoms as the 1st nearest neighbor in the middle of the slab is around 1.94-1.95Å, while the distance at the termination edges of the slab reduces to 1.80-1.94 Å during the relaxation. Similarly, the distance between N atoms is around 3.17 Å in the middle of the slab slightly decreasing to 3.14-3.15 Å at the termination edges. The calculated lattice constants of dichalcogenides in a bulk phase: $a$ = 3.13 Å and 3.14 Å for bulk $WS_2$ and $MoS_2$, respectively. These values are in good agreement with experimental data [58,59]. The calculated value of $c$ constant: $c$ = 12.44 Å and $c$ = 12.59 Å for $WS_2$ and $MoS_2$, respectively. The negligible differences between calculated and measured values ($c$=12.323 Å for $WS_2$ [58] and $c$ = 12.3 Å for $MoS_2$ [59]) could be explained by not using of empirical Grimme correction in the calculations. The reason for this is a poor SCF convergence of 2D nanoheterointerfaces if Grimme correction is in use.

Due to very close lattice constants $a$ of GaN substrate and both dichalcogenides the geometrical matching of the materials is good. The mismatch in the lattice constant during the stacking is 1.8%. The analysis of the atom positions after the relaxation reveals a certain alignment between Ga and S atoms and form a similar spatial arrangement as observed between Ga and N atoms in the bulk region similarly to Ref. [41]. For the constructed 2D nanoheterostructure the distances between Ga atoms in the middle of the slab are almost the same as in a perfect GaN slab and are equal to 3.18 Å, which indicates that the substrate thickness selected in the constructed model is sufficient. The relaxation of the atomic positions is larger at the termination edges of GaN substrate in comparison with the middle part of 2D heterostructure and are in the range from 3.02 Å to 3.21 Å. The relaxation of the interface of both $WS_2$@GaN and $MoS_2$@GaN heterostructures occurs similarly with almost identical atomic displacements. The difference between the heterostructures of $WS_2$@GaN and $MoS_2$@GaN is in the distance between GaN and $WS_2$ or $MoS_2$ layers, which slightly decreases from 4.25 Å to 4.24 Å for $WS_2$@GaN heterostructure and slightly increases from 4.25 Å to 4.27 Å for $MoS_2$@GaN heterostructure with the increase of the number of layers.

*Fig. 5* shows total densities of states (DOS) calculated for [0001]-oriented monolayered (1ML), bi-layered (2ML), and three-layered (3ML) $WS_2$ and $MoS_2$ nanofilms deposited atop of [1-100] oriented GaN NW. DOS calculated for free-standing GaN(1-100) slab to mimic the NW surface, as well as free-standing $WS_2$ and $MoS_2$ nanofilms are shown to elucidate changes in electronic structures induced by the *core-shell* interface formation. The top of the valence band (VB) of free-standing GaN(1-100) substrate (uppermost graph in *Figs. 5(a) and 5(b)*) is located at -6 eV with the band gap width of 3.9 eV. DOSes calculated for layered $WS_2$ nanofilm (three graphs situated at the bottom in *Fig. 5(a)*) exhibit the top of VB positioned at -6.7 eV for all free-standing $WS_2$ nanofilms. As the number of $WS_2$ layers increases, there is a certain decrease of the band gap width due to the quantum confinement effect. The calculated band gap for 1ML-, 2-ML, and 3ML-WS2 reduces in the following sequence: 2.6, 1.96, and 1.83 eV, respectively, with well-pronounced shift of the bottom of the conduction band (CB). Formation of $WS_2$@GaN interface results in further narrowing of the band gap width due to corresponding shift of both top of VB and bottom of the CB calculated for all 1ML-, 2ML-, and 3-ML-$WS_2$@GaN interfaces. The top of the VB calculated for all three $WS_2$@GaN interfaces under study remains mostly unchanged with respect to the number of nanofilm monolayers (-6.2 eV), while a more pronounced shift of the CB bottom leads to the gap of 1.32, 1.22, and 1.19 eV for 1ML-, 2ML-, and 3-ML-$WS_2$@GaN interfaces, respectively. DOS calculated for $MoS_2$@GaN interfaces (*Fig. 5(b)*) demonstrate a very similar behaviour of the band edges. Due to the quantum confinement effect the band gap of free-standing $MoS_2$ nanofilms increases with the decrease of the number of layers with the following sequence: 2.05, 2.23, and 2.72 eV for 3ML-, 2ML-, and 1ML-$MoS_2$, respectively. The shift of the VB top due to the formation of $MoS_2$@GaN interfaces leads to the further lowering of the band gap, while the CB bottom remains mainly unchanged. Band gaps calculated for 1ML-, 2ML-, and 3ML-$MoS_2$@GaN are 1.28, 1.19, and 1.17 eV, respectively.

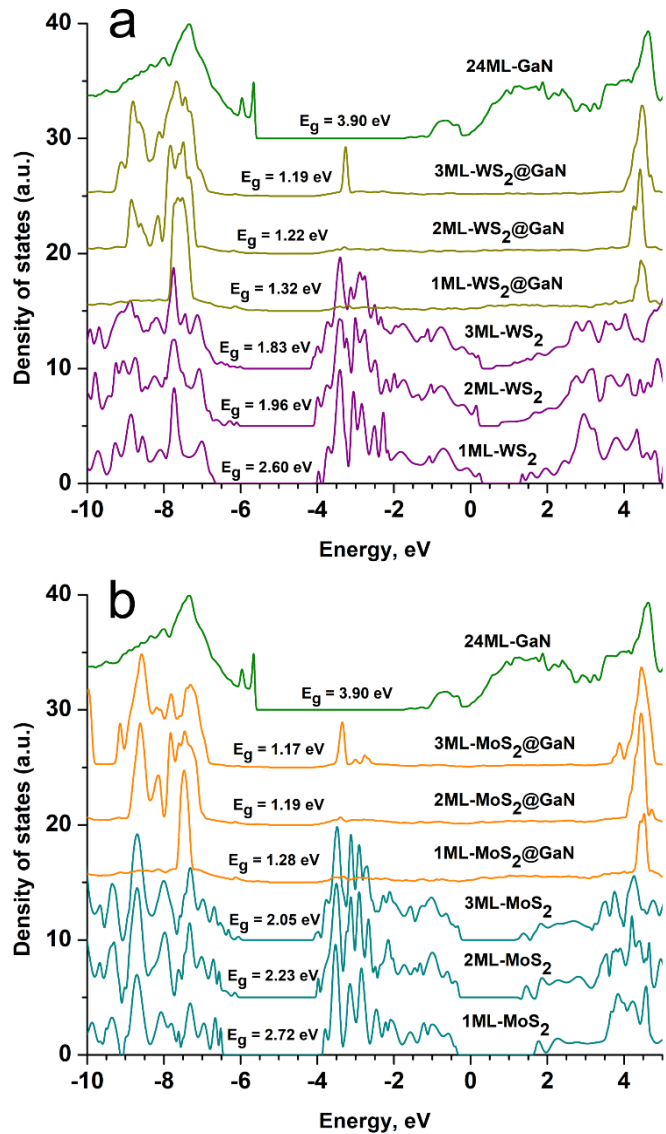

**Figure 5**. Density of states (DOS) as calculated for (a) $WS_2$@GaN and (b) $MoS_2$@GaN interfaces and their constituents. Zero corresponds to vacuum level. Each curve is normalized so that the maximum of each curve corresponds to 10 a.u.

According to the band edge diagram (*Fig. 6*) calculated for $WS_2$@GaN and $MoS_2$@GaN nanoheterostructures the free-standing $WS_2$ and $MoS_2$ monolayers have top of VB positioned below $O_2/H_2O$ redox potential, while CB bottom is properly positioned above standard hydrogen electrode (SHE). Increased number of both $WS_2$ and $MoS_2$ monolayers shifts VB top closer to oxygen redox potential. GaN substrate has VB top positioned at oxygen redox potential, while its CB bottom is located far above SHE. As soon as MoS$_2$ or WS$_2$ MLs are in contact with surface of GaN NW the CB bottom is positioned ~0.5 eV above SHE suitable for efficient hydrogen evolution within watersplitting reaction. VB top of both WS$_2$@GaN and MoS$_2$@GaN nanoheterostructures is located ~1 eV above O$_2$/H$_2$O redox level potentially making water splitting possible under red and near infrared irradiation. Thus, based on our calculations, WS$_2$@GaN and MoS$_2$@GaN *core-shell* NWs might be promising candidates for further investigation as photocatalysts for efficient hydrogen production from water.

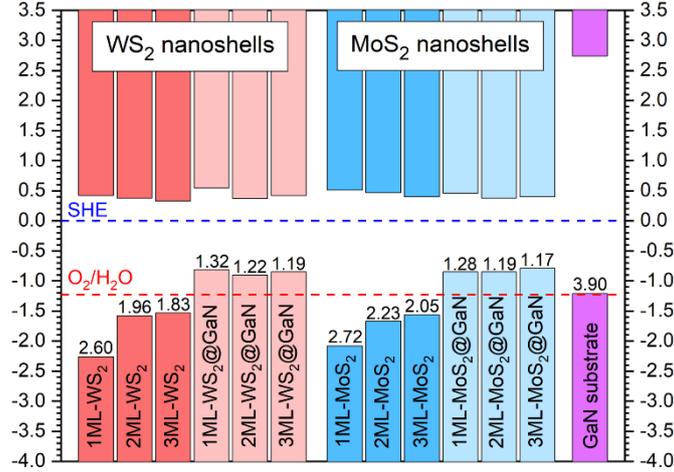

**Figure 6**. Schematic representation of the GaN-WS$_2$ and GaN-MoS$_2$ band edges positions with respect to the standard hydrogen electrode (SHE). Blue and red horizontal dashed lines correspond to the redox potentials for H$^+$/H$_2$ and O$_2$/H$_2$O dissociation, respectively.

## 5. Conclusions

In this work, we demonstrated growth of GaN-MoS$_2$ and GaN-WS$_2$ *core-shell* NWs via two different methods: (1) two-step process of sputter-deposition of a sacrificial transition metal oxide coating on GaN NWs followed by sulfurization; (2) pulsed laser deposition of few-layer MoS$_2$ or WS$_2$ on GaN NWs from the respective material targets. As-prepared nanostructures were characterized via scanning and transmission electron microscopies, X-ray diffraction, micro-Raman spectroscopy and X-ray photoelectron spectroscopy. GaN NWs are close to an ideal substrate for $MoS_2$ and $WS_2$ growth due to the very well matching crystal lattice and high chemical stability in the corrosive sulfur atmosphere. By changing the process parameters, it was possible to obtain either a smooth and uniform shell around the NWs, which is paramount for good electrical properties, or non-uniform island-like coating with increased surface roughness that could be beneficial for various energy applications. DFT calculations showed that such GaN-$MoS_2$ and GaN-$WS_2$ *core-shell* NWs might be promising candidates for further investigation as photocatalysts for efficient hydrogen production from water.

**Acknowledgements**

This research is funded by the Latvian Council of Science project "Core-shell nanowire heterostructures of Charge Density Wave materials for optoelectronic applications" No. lzp-2020/1-0261. E.B. was supported by the European Union's Horizon 2020 program, under Grant Agreement No. 856705 (ERA Chair "MATTER"). Institute of Solid State Physics, University of Latvia as the Center of Excellence has received funding from the European Union's Horizon 2020 Framework Programme H2020-WIDESPREAD-01-2016-2017-TeamingPhase2 under grant agreement No. 739508, project CAMART². The authors are grateful to Jevgenijs Gabrusenoks for micro-Raman measurements.

**Supplementary information**

Supplementary information is available and contains characterization data on pure GaN NWs, SEM images of GaN-MeS$_2$ NWs, TEM images of heterostructured NWs with various shell thickness, and projections of atomic structural models used in DFT calculations.

**References**

[1]     Novoselov, K. S., Mishchenko, A., Carvalho, A. & Castro Neto, A. H. 2D materials and van der Waals heterostructures. *Science (80-. ).* **353**, (2016)

[2]     Choi, W. *et al.* Recent development of two-dimensional transition metal dichalcogenides and their applications. *Mater. Today* **20**, 116–130 (2017)

[3]     Yazyev, O. V. & Kis, A. MoS2 and semiconductors in the flatland. *Mater. Today* **18**, 20–30 (2015)

[4]     Ponraj, J. S. *et al.* Photonics and optoelectronics of two-dimensional materials beyond graphene. *Nanotechnology* **27**, 462001 (2016)

[5]     Ali, Z. *et al.* Transition metal chalcogenide anodes for sodium storage. *Mater. Today* **35**, 131–167 (2020)

[6]     Di, J. *et al.* Ultrathin two-dimensional materials for photo- and electrocatalytic hydrogen evolution. *Mater. Today* **21**, 749–770 (2018)

[7]     Yang, S., Jiang, C. & Wei, S. Gas sensing in 2D materials. *Appl. Phys. Rev.* **4**, 021304 (2017)

[8]     Dasgupta, N. P. *et al.* 25th Anniversary Article: Semiconductor Nanowires - Synthesis, Characterization, and Applications. *Adv. Mater.* **26**, 2137–2184 (2014)

[9]     Jia, C., Lin, Z., Huang, Y. & Duan, X. Nanowire Electronics: From Nanoscale to Macroscale. *Chem. Rev.* **119**, 9074–9135 (2019)

[10]    Sun, Y., Sun, B., He, J. & Wang, C. Compositional and structural engineering of inorganic nanowires toward advanced properties and applications. *InfoMat* **1**, 496–524 (2019)

[11]    Lee, H. Y. & Kim, S. Nanowires for 2D material-based photonic and optoelectronic devices. *Nanophotonics* (2022) doi:10.1515/nanoph-2021-0800

[12]    Dan, Y. *et al.* Dramatic reduction of surface recombination by in situ surface passivation of silicon nanowires. *Nano Lett.* **11**, 2527–2532 (2011)

[13]  Mai, L., Tian, X., Xu, X., Chang, L. & Xu, L. Nanowire Electrodes for Electrochemical Energy Storage Devices. *Chem. Rev.* **114**, 11828–11862 (2014)

[14]  Chen, K., Shi, L., Zhang, Y. & Liu, Z. Scalable chemical-vapour-deposition growth of three-dimensional graphene materials towards energy-related applications. *Chem. Soc. Rev.* **47**, 3018–3036 (2018)

[15]  Butanovs, E. *et al.* Synthesis and characterization of GaN/ReS2, ZnS/ReS2 and ZnO/ReS2 core/shell nanowire heterostructures. *Appl. Surf. Sci.* **536**, 147841 (2021)

[16]  Chen, Z. *et al.* Core–shell MoO3–MoS2 Nanowires for Hydrogen Evolution: A Functional Design for Electrocatalytic Materials. *Nano Lett.* **11**, 4168–4175 (2011)

[17]  Choudhary, N. *et al.* High-Performance One-Body Core/Shell Nanowire Supercapacitor Enabled by Conformal Growth of Capacitive 2D WS2 Layers. *ACS Nano* **10**, 10726–10735 (2016)

[18]  Hu, C. *et al.* Hierarchical MoO3/SnS2 core-shell nanowires with enhanced electrochemical performance for lithium-ion batteries. *Phys. Chem. Chem. Phys.* **20**, 17171–17179 (2018)

[19]  Wang, X. *et al.* Boosting the stable sodium-ion storage performance by tailoring the 1D TiO2@ReS2 core-shell heterostructures. *Electrochim. Acta* **338**, 135695 (2020)

[20]  Kumar, K. S. *et al.* High-performance flexible asymmetric supercapacitor based on rGO anode and WO3/WS2 core/shell nanowire cathode. *Nanotechnology* **31**, 435405 (2020)

[21]  Tang, Z. R., Han, B., Han, C. & Xu, Y. J. One dimensional CdS based materials for artificial photoredox reactions. *J. Mater. Chem. A* **5**, 2387–2410 (2017)

[22]  Liu, S., Tang, Z. R., Sun, Y., Colmenares, J. C. & Xu, Y. J. One-dimension-based spatially ordered architectures for solar energy conversion. *Chem. Soc. Rev.* **44**, 5053–5075 (2015)

[23]  Li, J. Y., Yuan, L., Li, S. H., Tang, Z. R. & Xu, Y. J. One-dimensional copper-based heterostructures toward photo-driven reduction of CO2 to sustainable fuels and feedstocks. *J. Mater. Chem. A* **7**, 8676–8689 (2019)

[24]  Seo, B., Jeong, H. Y., Hong, S. Y., Zak, A. & Joo, S. H. Impact of a conductive oxide core in tungsten sulfide-based nanostructures on the hydrogen evolution reaction. *Chem. Commun.* **51**, 8334–8337 (2015)

[25] Khan, F., Idrees, M., Nguyen, C., Ahmad, I. & Amin, B. A first-principles study of electronic structure and photocatalytic performance of GaN–MX2 (M = Mo, W; X= S, Se) van der Waals heterostructures. *RSC Adv.* **10**, 24683–24690 (2020)

[26] Wang, J. *et al.* Thickness-Dependent Phase Stability and Electronic Properties of GaN Nanosheets and MoS2/GaN van der Waals Heterostructures. *J. Phys. Chem. C* **123**, 3861–3867 (2019)

[27] Kang, M. S., Lee, C.-H., Park, J. B., Yoo, H. & Yi, G.-C. Gallium nitride nanostructures for light-emitting diode applications. *Nano Energy* **1**, 391–400 (2012)

[28] Wang, G.-Z. *et al.* Tunable Photocatalytic Properties of GaN-Based Two-Dimensional Heterostructures. *Phys. status solidi* **255**, 1800133 (2018)

[29] Zhang, X. *et al.* Design and Integration of a Layered MoS2/GaN van der Waals Heterostructure for Wide Spectral Detection and Enhanced Photoresponse. *ACS Appl. Mater. Interfaces* **12**, 47721–47728 (2020)

[30] Zhuo, R. *et al.* High-performance self-powered deep ultraviolet photodetector based on MoS2/GaN p–n heterojunction. *J. Mater. Chem. C* **6**, 299–303 (2018)

[31] Jain, S. K. *et al.* Current Transport and Band Alignment Study of MoS2/GaN and MoS2/AlGaN Heterointerfaces for Broadband Photodetection Application. *ACS Appl. Electron. Mater.* **2**, 710–718 (2020)

[32] Goel, N. *et al.* A high-performance hydrogen sensor based on a reverse-biased MoS2/GaN heterojunction. *Nanotechnology* **30**, 314001 (2019)

[33] Zhang, Z., Qian, Q., Li, B. & Chen, K. J. Interface Engineering of Monolayer MoS2/GaN Hybrid Heterostructure: Modified Band Alignment for Photocatalytic Water Splitting Application by Nitridation Treatment. *ACS Appl. Mater. Interfaces* **10**, 17419–17426 (2018)

[34] Ghosh, D., Devi, P. & Kumar, P. Modified p-GaN Microwells with Vertically Aligned 2D-MoS2 for Enhanced Photoelectrochemical Water Splitting. *ACS Appl. Mater. Interfaces* **12**, 13797–13804 (2020)

[35] Liu, X. *et al.* Enhancing Photoresponsivity of Self-Aligned MoS2 Field-Effect Transistors by Piezo-Phototronic Effect from GaN Nanowires. *ACS Nano* **10**, 7451–7457 (2016)

[36] Yang, Z. *et al.* Performance Limits of the Self-Aligned Nanowire Top-Gated MoS2 Transistors. *Adv. Funct. Mater.* **27**, 1602250 (2017)

[37] Yang, G. *et al.* Chemical Vapor Deposition Growth of Vertical MoS2

Nanosheets on p-GaN Nanorods for Photodetector Application. *ACS Appl. Mater. Interfaces* **11**, 8453–8460 (2019)

[38] Wen, Y. *et al.* Direct growth of tungsten disulfide on gallium nitride and the photovoltaic characteristics of the heterojunctions. *Semicond. Sci. Technol.* **36**, 025016 (2021)

[39] Wu, L. *et al.* Enhancing the electrocatalytic activity of 2H-WS2 for hydrogen evolution via defect engineering. *Phys. Chem. Chem. Phys.* **21**, 6071–6079 (2019)

[40] Ji, Y. *et al.* Wrinkled-Surface-Induced Memristive Behavior of MoS2 Wrapped GaN Nanowires. *Adv. Electron. Mater.* **6**, 2000571 (2020)

[41] Zhou, B. *et al.* Gallium nitride nanowire as a linker of molybdenum sulfides and silicon for photoelectrocatalytic water splitting. *Nat. Commun.* **9**, 3856 (2018)

[42] Sung, D., Min, K.-A. & Hong, S. Investigation of atomic and electronic properties of 2D-MoS2/3D-GaN mixed-dimensional heterostructures. *Nanotechnology* **30**, 404002 (2019)

[43] Wan, Y. *et al.* Epitaxial Single-Layer MoS2 on GaN with Enhanced Valley Helicity. *Adv. Mater.* **30**, 1703888 (2018)

[44] Liao, J., Sa, B., Zhou, J., Ahuja, R. & Sun, Z. Design of High-Efficiency Visible-Light Photocatalysts for Water Splitting: MoS2/AlN(GaN) Heterostructures. *J. Phys. Chem. C* **118**, 17594–17599 (2014)

[45] Polyakov, B. *et al.* Unexpected Epitaxial Growth of a Few WS2 Layers on $\{1\bar{1}00\}$ Facets of ZnO Nanowires. *J. Phys. Chem. C* **120**, 21451–21459 (2016)

[46] http://www.crystal.unito.it/.

[47] Dovesi, R. *et al.* The CRYSTAL code, 1976–2020 and beyond, a long story. *J. Chem. Phys.* **152**, 204111 (2020)

[48] Heyd, J., Scuseria, G. E. & Ernzerhof, M. Erratum: "Hybrid functionals based on a screened Coulomb potential" [J. Chem. Phys. 118, 8207 (2003)]. *J. Chem. Phys.* **124**, 219906 (2006)

[49] Pandey, R., Jaffe, J. E. & Harrison, N. M. Ab initio study of high pressure phase transition in GaN. *J. Phys. Chem. Solids* **55**, 1357–1361 (1994)

[50] Gatti, C., Saunders, V. R. & Roetti, C. Crystal field effects on the topological properties of the electron density in molecular crystals: The case of urea. *J. Chem. Phys.* **101**, 10686–10696 (1994)

[51] Corà, F., Patel, A., Harrison, N. M., Dovesi, R. & Catlow, C. R. A. An ab Initio

Hartree−Fock Study of the Cubic and Tetragonal Phases of Bulk Tungsten Trioxide. *J. Am. Chem. Soc.* **118**, 12174–12182 (1996)

[52] Corà, F., Patel, A., Harrison, N. M., Roetti, C. & Catlow, C. R. A. An ab initio Hartree–Fock study of α-MoO3. *J. Mater. Chem.* **7**, 959–967 (1997)

[53] Lichanot, A., Aprà, E. & Dovesi, R. Quantum Mechnical Hartree-Fock Study of the Elastic Properties of Li2S and Na2S. *Phys. status solidi* **177**, 157–163 (1993)

[54] Monkhorst, H. J. & Pack, J. D. Special points for Brillouin-zone integrations. *Phys. Rev. B* **13**, 5188–5192 (1976)

[55] Grimme, S., Antony, J., Ehrlich, S. & Krieg, H. A consistent and accurate ab initio parametrization of density functional dispersion correction (DFT-D) for the 94 elements H-Pu. *J. Chem. Phys.* **132**, 154104 (2010)

[56] Yun, Q. *et al.* Layered Transition Metal Dichalcogenide-Based Nanomaterials for Electrochemical Energy Storage. *Adv. Mater.* **32**, 1903826 (2020)

[57] Tonndorf, P. *et al.* Photoluminescence emission and Raman response of monolayer MoS2, MoSe2, and WSe2. *Opt. Express* **21**, 4908 (2013)

[58] Schutte, W. J., De Boer, J. L. & Jellinek, F. Crystal structures of tungsten disulfide and diselenide. *J. Solid State Chem.* **70**, 207–209 (1987)

[59] Wakabayashi, N., Smith, H. G. & Nicklow, R. M. Lattice dynamics of hexagonal MoS2 studied by neutron scattering. *Phys. Rev. B* **12**, 659–663 (1975)